\documentclass[english,notitlepage]{revtex4-1}
\usepackage[T1]{fontenc}
\usepackage[latin9]{inputenc}
\usepackage{color}
\usepackage{amsmath}
\usepackage{amssymb}
\usepackage{graphicx}

\makeatletter
\usepackage{babel}

\newcommand{\E}{{\mbox{\scriptsize \sc E}}}
\newcommand{\I}{{\mbox{\scriptsize \sc I}}}
\newcommand{\X}{{\mbox{\scriptsize \sc X}}}
\newcommand{\Y}{{\mbox{\scriptsize \sc Y}}}
\newcommand{\T}{{\mbox{\scriptsize \sc T}}}

\newcommand{\inS}{{\mbox{\scriptsize in}}}
\newcommand{\out}{{\mbox{\scriptsize out}}}
\newcommand{\thr}{{\mbox{\scriptsize th}}}
\newcommand{\preM}{{\mbox{\scriptsize pre}}}
\newcommand{\res}{{\mbox{\scriptsize res}}}
\newcommand{\refR}{{\mbox{\scriptsize ref}}}
\newcommand{\m}{{\mbox{\scriptsize m}}}
\newcommand{\f}{{\mbox{\scriptsize f}}}
\newcommand{\test}{{\mbox{\scriptsize test}}}
\newcommand{\s}{{\mbox{\scriptsize s}}}
\newcommand{\eff}{{\mbox{\scriptsize eff}}}
\newcommand{\run}{{\mbox{\scriptsize run}}}

\makeatother

\usepackage{babel}
\begin{document}

\title{Training dynamically balanced excitatory-inhibitory networks}

\author{Alessandro Ingrosso}

\author{L.F. Abbott}

\affiliation{Zuckerman Mind, Brain, Behavior Institute, Columbia University, New
York, NY 10027}
\begin{abstract}
The construction of biologically plausible models of neural circuits
is crucial for understanding the computational properties of the nervous
system. Constructing functional networks composed of separate excitatory
and inhibitory neurons obeying Dale's law presents a number of challenges.
We show how a target-based approach, when combined with a fast online
constrained optimization technique, is capable of building functional
models of rate and spiking recurrent neural networks in which excitation
and inhibition are balanced. Balanced networks can be trained to produce
complicated temporal patterns and to solve input-output tasks while
retaining biologically desirable features such as Dale's law and response variability. 
\end{abstract}
\maketitle

\section*{Introduction}

Cortical neurons typically require only a small fraction of their
thousands of excitatory inputs to reach firing threshold. This suggests
an overabundance of excitation that must be balanced by inhibition
to keep neurons within their functional operating ranges. An interesting
suggestion is that this balance does not require fine-tuning of synaptic
strengths, what we will call parametric balance, but rather occurs
dynamically \cite{VanVreeswijkChaosScience,VanVreeswijkChaoticBalancedStateNeuralComputation,RenartAsynchronous,KadmonSompolinsky,HarishHanselAsynchronous,mastrogiuseppeeinetworks,TsodyksStateSwitching,BrunelDynamicsSparsely}.
Dynamically balanced neural network models were originally introduced
to account for the high variability of neural activity. Variants of
balanced networks have since been used to model response selectivity
\cite{HanselMechanismOrientation,PehlevanSelectivity} and associative memory \cite{RubinBalanced},
but a general approach to task learning in these models has not previously
been developed. The challenge is that the mean activity of dynamically
balanced network models is constrained. To maintain dynamic balance,
a learning scheme must flexibly alter network dynamics while respecting
this constraint, otherwise the network will transition to a parametrically
balanced regime. In addition, balanced networks display a reduced
mean response to stimuli and, as a consequence, suppressing chaos
in these networks, a critical step in learning, requires some care.
We present approaches for training networks while both suppressing
chaos and retaining dynamic balance.

In addition to the issues with balancing outlined above, training
networks with sign-constrained weights presents some technical challenges.
Batch approaches to learning can handle sign constraints quite efficiently,
but batch training of recurrent networks often leads to instabilities
during testing, even when the training error is small \cite{sussillo2009generating,JaegerTutorial}.
The use of an online strategy is critical to quench spontaneous chaotic
fluctuations during training and to assure stability of the trained
dynamics. These requirements demand fast learning algorithms capable
of adjusting weights as the network is running. In previous work \cite{sussillo2009generating,LajeBuonomanoTamingChaos,Vincent-LamarreDrivingReservoir},
this was achieved by using a recursive least squares (RLS) algorithm
that has the favorable feature of constraining network dynamics while
permitting fluctuations during training that are critical for post-training
stability. Unfortunately, when sign-constraints are imposed, standard
online training procedures, including RLS, are no longer viable. Here,
we developed a fast sign-constrained online method that proves effective
at training both rate and spiking balanced network models.

\section*{Results}

\subsection*{Dynamically and Parametrically Balanced Networks}

The networks we consider are composed of either spiking neurons interacting
via synaptic currents or so-called rate units. A task is generally
specified by a set of desired output signals $F_{k}^{\out}\left(t\right)$,
for $k=1,2,\ldots K_{\out}$ that are read out through channels $z_{k}$.
These signals can either be autonomously generated by the network
or arise in response to $K_{\inS}$ external inputs $F_{k}^{\inS}$$\left(t\right)$
entering the network through input weight vectors $\boldsymbol{w}_{k}^{\inS}$.
The input weights are generally chosen randomly and not subject to
learning, whereas the readout weights, which are not sign-constrained,
are trained using RLS. In rate models, $z_{k}=\boldsymbol{w}_{k}^{\out}\cdot\phi\left(\boldsymbol{x}\right)$,
where $\phi(x)$ is the rate activity for a unit with total input
$x$. The equations of the $N$ units of the network, for $i=1,2,\ldots,N$,
are 
\begin{equation}
\tau\frac{dx_{i}}{dt}=-x_{i}+\sum_{j=1}^{N}J_{ij}\phi\left(x_{j}\right)+\sum_{k=1}^{K_{\inS}}w_{ik}^{\inS}F_{k}^{\inS}+I_{i}\label{eq:rate_eq}
\end{equation}
where $I\in\left\{ I_{\E},I_{\I}\right\} $ is a vector of constant and uniform external
currents into the E and I populations, and $w_{:k}^{\inS}$ are the
weight vectors for each of the $K_{\inS}$ input channels. We employ
a variety of activation functions, e.g.\ halftanh ($\phi\left(x\right)=\theta(x)\tanh\left(x\right)$),
sigmoid ($\phi\left(x\right)=1/\left(1+\exp\left(-x\right)\right)$)
or ReLU ($\phi\left(x\right)=\theta\left(x\right)x$), where $\theta$
is the Heaviside step function ($\theta\left(x\right)=1$ when $x>0$
and $0$ otherwise).

For the spiking networks, we use leaky integrate-and-fire (LIF) dynamics
(although good performances can be achieved with other neuronal models)
of the form 
\begin{eqnarray}
\tau_{\m}\frac{dV_{i}}{dt} & = & -V_{i}+\sum_{j=1}^{N}J_{ij}s_{j}+\sum_{k=1}^{K_{\inS}}w_{ik}^{\inS}F_{k}^{\inS}+I_{i}\label{eq:if_voltage}\\
\tau_{\s}\frac{ds_{i}}{dt} & = & -s_{i}+\tau_{\s}\sum_{t_{i}^{\f}<t}\delta\left(t-t_{i}^{\f}\right)\label{eq:if_synapse}
\end{eqnarray}
where $\tau_{\m}$ is the membrane time constant ($\tau_{\m}=20$
ms in all simulations) and $t_{i}^{\f}$ is a list of the times when
neuron $i$ fired. When $V_{i}\left(t\right)$ reaches the spiking
threshold $V_{\thr}$ (usually set to $1$) a spike is emitted and
the voltage $V_{i}$ is reset to $V_{\res}$ and kept constant for
a period of time equal to the refractory period $\tau_{\refR}$. We
typically take either $\tau_{\refR}=2$ ms or no refractoriness ($\tau_{\refR}=0$), and $\tau_{\s}=50$ ms or $\tau_{\s}=100$ ms. The readouts for spiking networks are given by $z_{k}=\boldsymbol{w}_{k}^{\out}\cdot\boldsymbol{s}$.

For networks with distinct excitatory and inhibitory neurons, the
connection matrix $J$ in equations \eqref{eq:rate_eq} and~\eqref{eq:if_voltage}
is divided into 4 blocks, $J_{\E\E}$, $J_{\E\I}$, $J_{\I\E}$ and
$J_{\I\I}$, where the first and second subscripts denote the type
of post- and presynaptic neurons, respectively. We are interested
in training networks that are in a so-called balanced regime for which
the strengths of individual synapses are of order $1/\sqrt{N_{\preM}}$,
where $N_{\preM}$ is the number or pre-synaptic neurons, which in our
case is $\mathcal{O}\left(N\right)$.  Because of this and because we focus on cases with equal numbers of E and I neurons (although other choices yield similar results), we will not distinguish between orders of $\sqrt{N_{\preM}}$ and $\sqrt{N}$ and stick to the latter notation. Sign-constrained weights combined with non-negative firing rates implies that the sum of the inputs over all excitatory or all the inhibitory synapses is
of order $N/\sqrt{N}=\sqrt{N}$. If uncanceled, this would introduce an extremely large term into equations \eqref{eq:rate_eq} and~\eqref{eq:if_voltage} that would lead to either nearly silenced or
saturated dynamics. The solution to this problem is to introduce a
compensating large constant input, $I_{i}=[\alpha_{\E}\sqrt{N},\alpha_{\I}\sqrt{N}]$
that is also of order $\sqrt{N}$ ($\alpha$ is of order $1$).
Cancellation of all the terms in equations \eqref{eq:rate_eq} and~\eqref{eq:if_voltage}
of order $\sqrt{N}$ then leads to the constraint 
\begin{equation}
J^{\eff}\left(\begin{array}{c}
r_{\E}\\
r_{\I}
\end{array}\right)\sim-\left(\begin{array}{c}
\alpha_{\E}\\
\alpha_{\I}
\end{array}\right)\quad\mbox{where}\quad J^{\eff}=\sqrt{N}\left(\begin{array}{cc}
\overline{J}_{\E\E} & \overline{J}_{\E\I}\\
\overline{J}_{\I\E} & \overline{J}_{\I\I}
\end{array}\right)\,,\label{balanceEq}
\end{equation}
and the symbol $\sim$ implies equality to within a discrepancy of
order $1/\sqrt{N}$. In this equation, $\overline{J}_{\X\Y}$
is the average of the elements in the submatrix $J_{\X\Y}$, and $r_{\E}$
and $r_{\I}$ are the average firing rates of the excitatory and inhibitory
populations. In a dynamically balanced network, the means of the excitatory
and inhibitory populations are both constrained by equation~\eqref{balanceEq}.  As we will show, there are different regimes for trained parametrically balanced networks.  When $I = 0$ or of order 1, equation~\eqref{balanceEq} forces extremely small rates unless the determinant of $J^{\eff}$ is of order $1/\sqrt{N}$, resulting in a parametrically balanced situation.  If $I$ is of order $\sqrt{N}$, we find a different kind of parametric balance that involves another form of fine-tuned cancelation, as discussed below.

In our experience, many learning schemes result in connection matrices
that realize a parametric rather than dynamic balance \cite{RubinBalanced}.
This comes about even if the initial connectivity $J$ has a $J^{\eff}$
with determinant of order 1\@. One common way for this to occur is if learning imposes a symmetry on the $J$ matrix so that the excitatory
and inhibitory mean weight values are proportional to each other.
We now show that an online learning scheme, combined with the appropriate
regularization, can construct dynamically balanced models that solve
a variety of tasks.

\subsection*{Full-Force in E/I Networks}

\begin{figure}
\centering{}\includegraphics[width=0.6\textwidth]{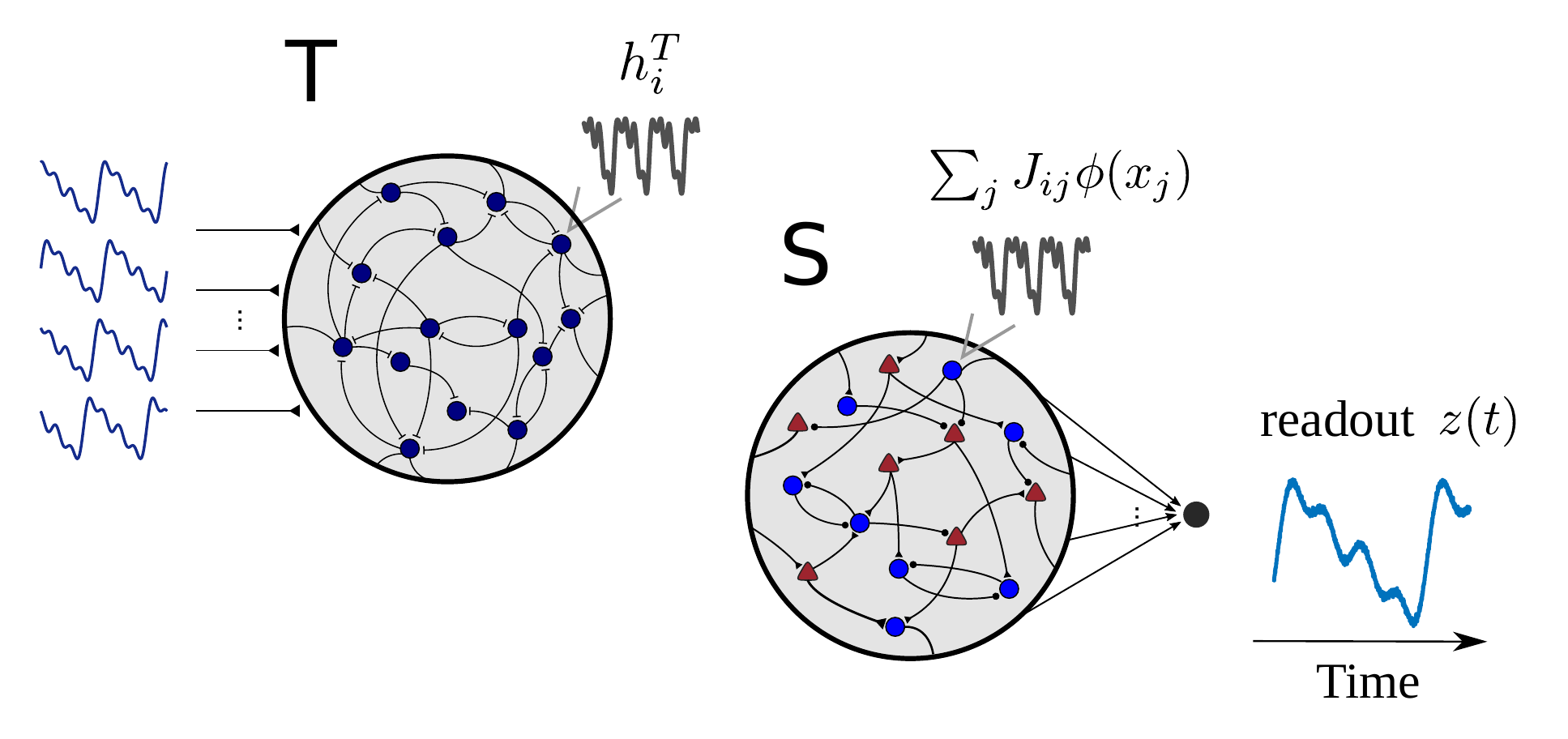}\caption{\label{fig:Schematic_target_based} \emph{Schematic of the target-based
method}. Target currents $h_{i}^{\T}(t)$ are produced by a balanced
teacher network (T, Left) driven by the desired output. The student
network (Right) is trained to reproduce the target currents autonomously.
We train the recurrent weights of both the excitatory (E) and inhibitory
(I) populations, together with the connections between them. A linear decoder $\boldsymbol{w}^{\out}$ is trained
with a standard online method (RLS) to reproduce the prescribed output
target from a readout of the neurons in the network.}
\end{figure}

We build upon a previously developed target-based approach for training
rate and spiking networks \cite{abbott2016building,depasquale2016using,DePasqualeFullForce}
(Fig.\ \ref{fig:Schematic_target_based}). In this scheme, a teacher
network (T), which in the cases we consider is an E/I rate model,
is driven by the desired output signals $F_{k}^{\out}$$\left(t\right)$.
This is done by adding a term $\sum_{k=1}^{K_{\out}}w_{ik}^{T}F_{k}^{\out}$
to equation~\eqref{eq:rate_eq} with random weights $w_{ik}^{\T}$
(we use superscript T to denote quantities associated with the teacher
network). We then extract a set of target currents, 
\begin{equation}
h_{i}^{\T}\!\left(t\right)=\sum_{j=1}^{N}J_{ij}^{\T}\phi\left(x_{j}^{\T}\left(t\right)\right)+\sum_{k=1}^{K_{\out}}w_{ik}^{\T}F_{k}^{\out}\left(t\right) \, ,
\end{equation}
from the teacher network. The full recurrent synaptic matrix $J$ of the
network we are training (called the student network; variables without
superscripts T are associated with the student network) is then trained
to generate these target currents autonomously without any driving
input. Specifically, for each neuron the training goal is to minimize
the cost function, for a run of duration $t_{\run}$, $E = \sum_i E_i$ with 
\begin{equation}
E_{i}=\frac{1}{t_{\run}}\int_{0}^{t_{\run}} dt\Big(h_{i}^{\T}\!\left(t\right)-\sum_{j=1}^{N}J_{ij}\phi\left(x_{j}(t)\right)\Big)^{2}+\alpha R_{i}\, .
\label{eq:loss_no_penalties}
\end{equation}
$R_{i}$ is a regularization term to be discussed below. In our case, the expression in equation~\eqref{eq:loss_no_penalties} is minimized subject to sign constraints on the elements of the matrix
$J$.

In the original full-FORCE scheme \cite{abbott2016building,depasquale2016using}, the cost~\eqref{eq:loss_no_penalties} is minimized using RLS but, as discussed above, this is not
a viable procedure when sign constraints are imposed. Instead,
we use bounded constrained coordinate descent (BCD) \cite{wright2015coordinate},
which proves to be a fast and reliable strategy for training both
rate and spiking models with sign constrained weights (Methods).
The resulting learning algorithm is fast enough to effectively clamp
the network dynamics close to the desired trajectory during training,
suppressing chaos and assuring stability.

\subsection*{Training Dynamically Balanced Networks}

\begin{figure}
\centering{}\includegraphics[width=0.8\textwidth]{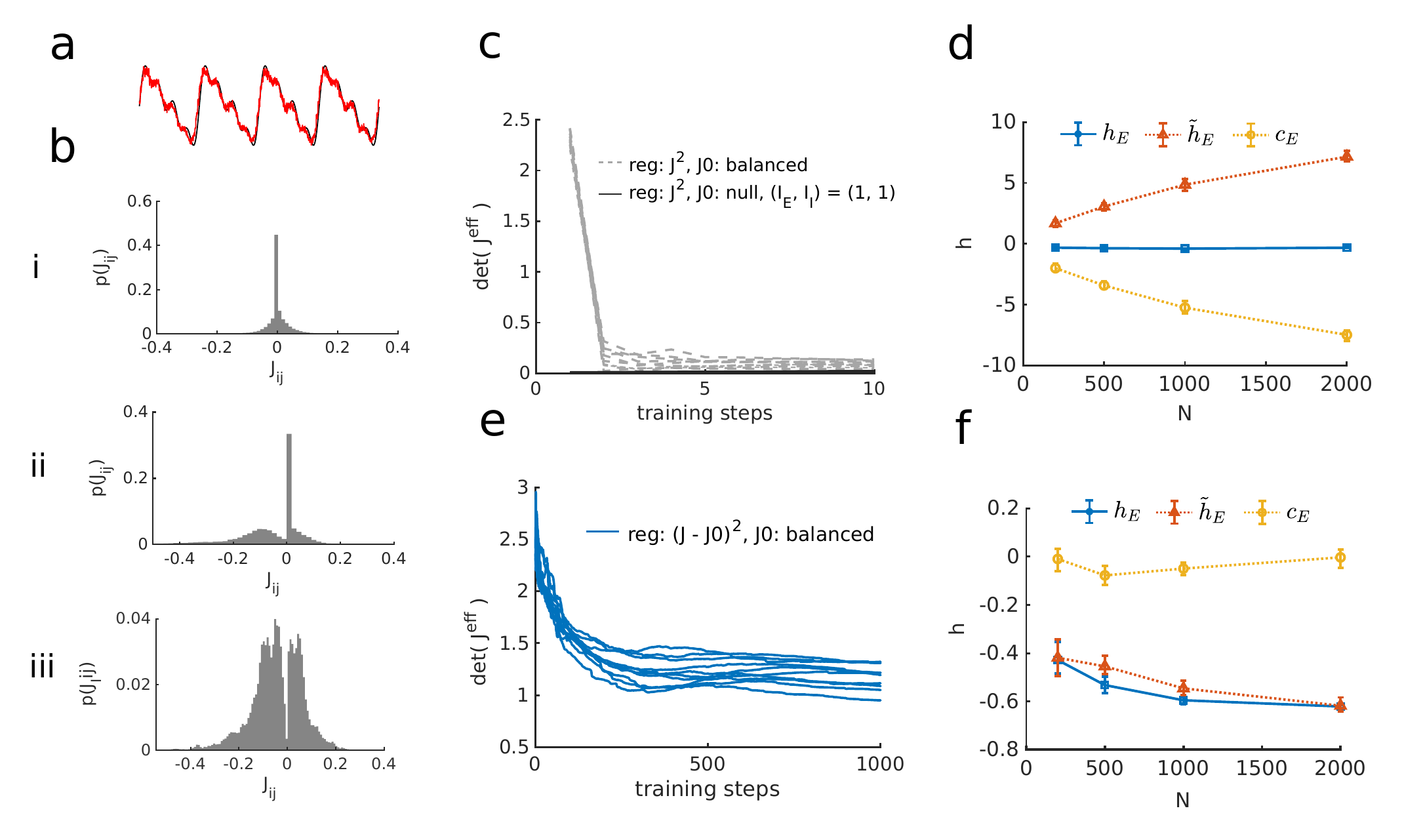}\caption{\emph{\label{fig:balanced_networks_weights_and_dets}Trained balanced
networks}. \textbf{a) }Target output $F^{out}$ (in black) for all 
the networks in this figure. Red curve is an example readout $z(t)$
from a trained spiking network of $N=200$ units. \textbf{b}) Histogram
of recurrent weights in three prototypical trained rate networks ($N=300$,
$\phi=$ halftanh): \textbf{i)} Zero external current ($I_{\E}=I_{\I}=0$)
and L2 regularization; \textbf{ii) }$[I_{\E},\:I_{\I}]=(0.3\sqrt{N},0.4\sqrt{N})$ and L2 regularization; \textbf{iii) }balanced initialization and J0 regularization, external currents as in ii. Regularization
parameter $\alpha=1.0$ in all three cases. \textbf{c})
Time course of the determinant of the effective matrix $J^{\eff}$
during training of spiking networks of size $N=200$ for $I$ of order $\sqrt{N}$ (grey dashed lines)
and $I$ of order 1 (black line on horizontal axis).  Both cases use L2 regularization. \textbf{d})
The full excitatory current and its two defined components (equation \eqref{eqnDefComps}) as a function of N for a parametrically balanced network performing the task in panel a\@.  \textbf{e})
Time course of the determinant of the effective matrix $J^{\eff}$
during training of spiking networks of size $N=200$ for $I$ of order $\sqrt{N}$ and J0 regularization. \textbf{f})
The full excitatory current and its two defined components (equation \eqref{eqnDefComps}) as a function of N for a dynamically balanced network performing the task in panel a\@.  Results in c-f are from ten different initializations of $J^0$ or $J^{\T}$.}
\end{figure}

For a given task, the distribution of synaptic weights after training
depends on a variety of factors including the initial value of the
$J$ matrix, which we call $J^{0}$, the choice of regularizer, and
whether the network is tonically driven by large constant external
current ($I$ in equation~\eqref{eq:rate_eq} and~\eqref{eq:if_voltage}). We begin by considering
a task in which the network must autonomously (meaning with time-independent
input) generate the periodic output shown in Fig.\ \ref{fig:balanced_networks_weights_and_dets}a. When no constant external current is present ($I=0$), equation~\eqref{balanceEq}
requires a parametric balance for any appreciable activity to exist
in the network. The resulting parametrically balanced network can
perform the task, but we find that an extensive fraction of synaptic
weights are set to zero by the training algorithm, so that the resulting
networks display a connection probability $\sim0.5$ and a symmetric
weight distribution (Fig.\ \ref{fig:balanced_networks_weights_and_dets}b,i). In the presence of constant external currents of order $\sqrt{N}$, the network has the potential to be dynamically balanced, but we find that, with a commonly used L2 weight regularization
($R_{i} =\sum_{j}J_{ij}^{2}$), the network also goes into a parametrically balanced configuration, though of a different form. This occurs regardless of the structure of the teacher network or the value of $\det{J^{\eff}}$ for the initial weights $J^{0}$. In this case, the weight distribution typically shows an extensive number of zero weights and a distribution of excitatory synapses that
is approximately Gaussian but cut-off at zero (Fig.\ \ref{fig:balanced_networks_weights_and_dets}b,ii). The determinant of $J^{\eff}$ is small but, unlike the case with zero external current, it is not of order $1/\sqrt{N}$ (Fig.\ \ref{fig:balanced_networks_weights_and_dets}c).  

To understand the nature of the parametric balance exhibited by networks with $I$ or order $\sqrt{N}$ trained with an L2 regularizer, we averaged the total input to the excitatory neurons and divided the result into two pieces,
\begin{equation}
h_{\E}=\left\langle\frac{1}{N_{\E}}\sum_{i\in \E}\sum_{j}J_{ij}r_{j}\right\rangle+I_{\E}=\underbrace{N_{\E}J_{\E\E}\langle r_{\E}\rangle+N_{\I}J_{\E\I}\langle r_{I}\rangle+I_{\E}}_{\tilde{h}_{\E}}+\underbrace{\frac{1}{N_{\E}}\sum_{i\in \E}\sum_{j}\delta J_{ij}\langle r_{j}\rangle}_{c_{\E}}\, ,
\label{eqnDefComps}
\end{equation}
where we have introduced $\delta J$ as the connection matrix with the block mean values removed, and brackets denote time averages.  In a standard dynamically balanced case, both $\tilde h_{\E}$ and $c_{\E}$ are of order 1, as is $h_{\E}$.  In contrast, for the network trained with the L2 regularizer, both $\tilde h_{\E}$ and $c_{\E}$ are of order $\sqrt{N}$, but they cancel to produce $h_{\E}$ of order 1 (Fig.\ \ref{fig:balanced_networks_weights_and_dets}d).  This cancellation is due to a fine-tuning of $J$ that arises during the learning process.

These results illustrate that dynamically balanced networks do not arise naturally from learning, even if the teacher network and the initial weight matrix of the student network are configured to be
dynamically balanced and $I$ is of order $\sqrt{N}$. The learning algorithm with L2 regularization tends to push the weight matrix to a parametrically balanced regime. We found a simple way to prevent this: choose $J^{0}$ to satisfy the dynamically balanced condition (stable solution to equation \eqref{balanceEq} with order 1 rates) and use regularization to keep $J$ from straying too far from $J^{0}$. The regularization that does this still uses an L2 norm, but on the difference between $J$ and $J^{0}$ rather than on the magnitude of $J$. Specifically, we define what we call the J0 regularizer by $R_{i}=\sum_{j}(J_{ij}-J_{ij}^{0})^{2}$.  With this regularizer, the weights after training display a Gaussian-like distribution (Fig.\ \ref{fig:balanced_networks_weights_and_dets}b,iii), block-wise average weights scaling as $1/\sqrt{N}$ and a $J^{\eff}$ determinant of order 1 (Fig.\ \ref{fig:balanced_networks_weights_and_dets}e).  Furthermore, the total current $h_{\E}$ and the two components we have introduced, $\tilde h_{\E}$ and $c_{\E}$, are all of order 1 (Fig.\ \ref{fig:balanced_networks_weights_and_dets}f).  Thus, dynamically balanced networks trained in this way, even when they are fairly small, have average activities and currents in agreement with what is expected from a dynamically balanced regime (however, deviations  from the usual balance constraints are slightly larger than the deviations seen for randomly connected networks due to weight-rate correlations generated by learning).

\begin{figure}
\centering{}\includegraphics[width=0.6\textwidth]{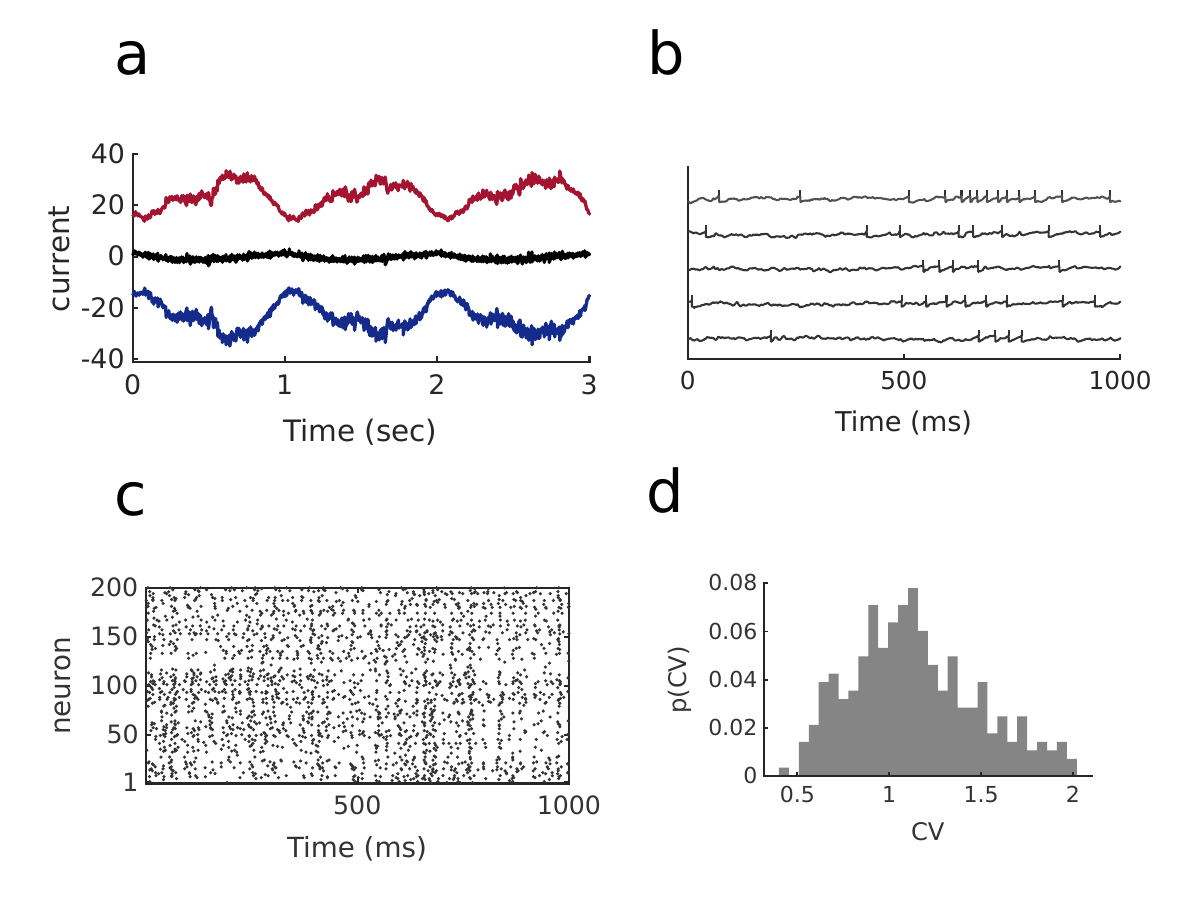}\caption{\label{fig:balanced_networks_dynamics}\emph{Dynamics in dynamically balanced trained spiking
networks.} \textbf{a}) Input currents onto a neuron in a spiking network
trained to produce a superposition of 4 sine waves as in Fig.\ \ref{fig:balanced_networks_weights_and_dets}a. Red curve: total excitatory current $h_{\E}+I_{\E}$;
Blue curve: inhibitory synaptic current $h_{\I}$; black curve: total
current $h$. \textbf{b}) Voltage traces of $5$ sample units the network with random fast synaptic currents (time constant 2 ms). \textbf{c}) Spike raster of $200$ neurons for the network in
b. \textbf{d}) Histogram of the coefficient of variation of interspike intervals
across neurons for the network in b. }
\end{figure}

We can use BCD and J0 regularization to train dynamically balanced spiking
networks as well (Fig.\ \ref{fig:balanced_networks_dynamics}). One common consequence of employing long synaptic time-scales is that a bursty spiking behavior emerges. The level of burstiness in trained networks can be varied by means of the $\omega_{h}$ parameter, that scales the intensity of the learned currents, generated by the slow synapses, with respect to the contribution provided by the random synapses with a fast time-constant (Methods).  The irregularity of spiking in trained networks depends on the amplitude of the current
fluctuations. To generate irregular spiking (Fig.\ \ref{fig:balanced_networks_dynamics}b-d),
we included random untrained fast-synapses (with synaptic time constant 2 ms;
see \cite{depasquale2016using}) and an average excess of inhibition.
The level of spiking irregularity can be quantified by computing the
distribution of coefficient of variations (CV) of interspike intervals
across the neurons of the network (\ref{fig:balanced_networks_dynamics}d). The average CV$\thickapprox1$.

\subsubsection*{Perturbations in trained balanced networks}

\begin{figure}
\centering{}\includegraphics[width=0.7\textwidth]{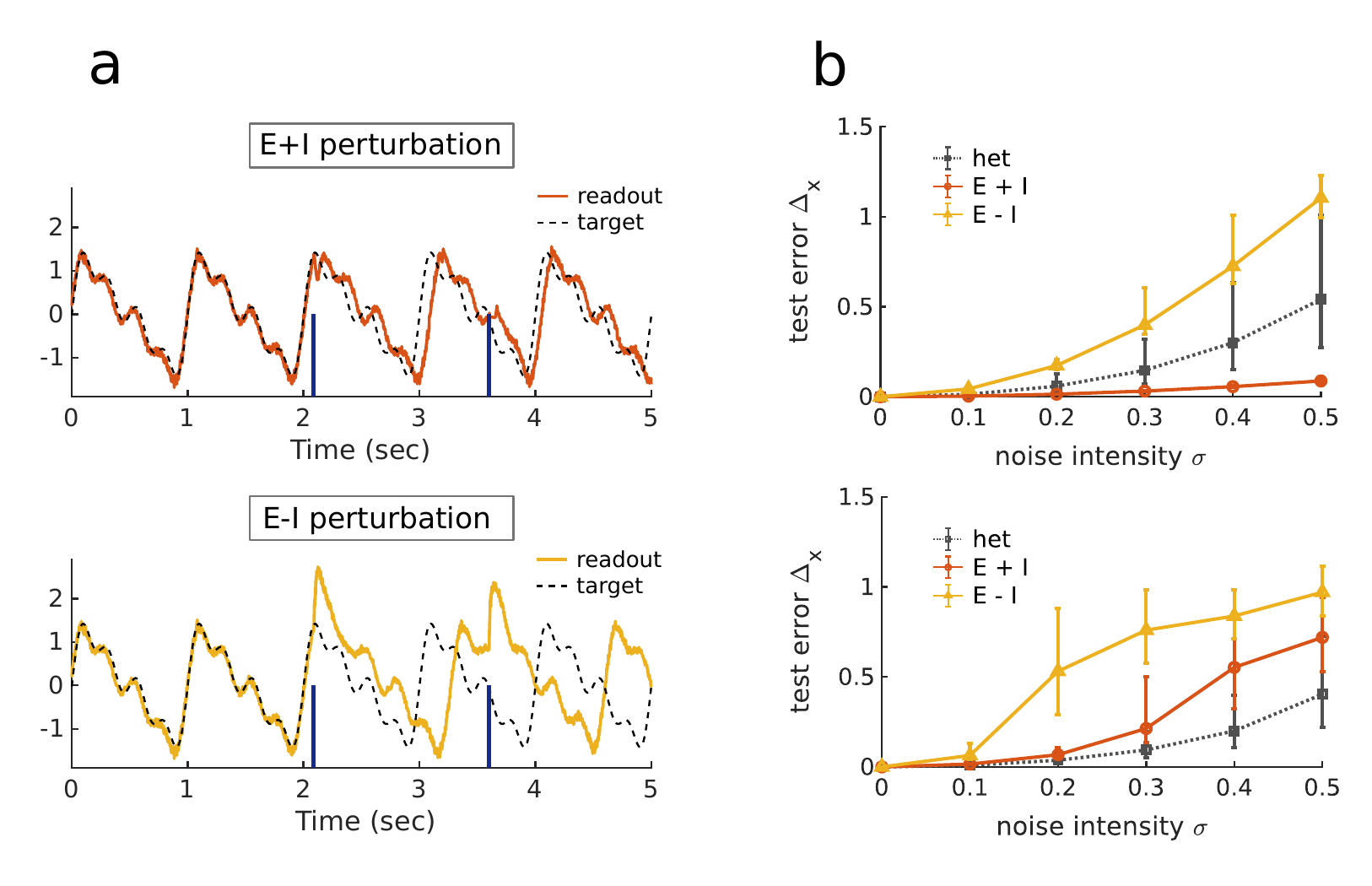}\caption{\label{fig:balance_perturbations}\emph{Response to perturbations
in trained balanced networks}. \textbf{a}) An E/I spiking network
of size $N=200$ trained on an oscillation task receives strong
input pulses at random times (dark blue vertical lines), either in
the E+I direction (top) or in the E-I direction (bottom). \textbf{b})
Median test error of two types of rate networks of size $N=200$ trained
to produce the same output signal as in a. Errorbars indicate
$25\%$ and $75\%$ percentiles over $100$ networks and $50$ realizations
of input white noise with intensity $\sigma$. The networks are
driven either by $N$ independent white noise inputs (black curve, legend: het) or by a single common white noise input in the E+I (red curve, legend E+I) or E-I (yellow curve, legend E-I) direction. Top: dynamically balanced network; bottom: parametrically balanced network with zero external
input. Halftanh activation function, see equation \eqref{eqnDelta} for the definition of $\Delta_x$.}
\end{figure}

Balanced networks trained on autonomous oscillation tasks can suppress homogeneous perturbations in a way similar to
the decorrelation effect mediated by the strong inhibitory feedback in such networks
\cite{RenartAsynchronous,HeliasDecorrelation}. As an example, we
consider spiking networks trained to reproduce autonomously the periodic
signal shown in Fig.\ \ref{fig:balanced_networks_weights_and_dets}a. We constructed both dynamically and parametrically balanced examples
of these networks and perturbed them at random times with 10 ms
duration current pulses. These pulses come in two types, either identical
for all neurons, or identical in magnitude but opposite in sign for
excitatory and inhibitory neurons, with positive input to the excitatory
neurons. We call these E+I and E-I perturbations, respectively. Balanced networks
generally exhibit a strong resilience to E+I perturbations  (Fig.\ \ref{fig:balance_perturbations}a, top) compared to external pulses in the E-I direction (Fig.\ \ref{fig:balance_perturbations}a, bottom). The latter produce a longer lasting transient and a subsequent larger phase shift in the network output. This response
to temporary imbalance in the collective activity of the E and I populations
is reminiscent of balance-amplified transients, previously described
by a linear theory \cite{MurphyMillerBalancedAmplification}.

The role of inhibitory feedback is also apparent when a rate network
is trained to produce the same rhythmic behavior. In this case, we perturbed the network with ongoing noise rather than with a transient.  Homogeneous E+I input disturbances are cancelled by strong inhibitory recurrence in dynamically (Fig.\ \ref{fig:balance_perturbations}b, top) but not in parametrically (Fig.\ \ref{fig:balance_perturbations}b, bottom) balanced networks.  E-I perturbations produce the strongest effect, and random heterogeneous perturbations produce similar effects in both networks, which are intermediate between E+I and E-I perturbations in the dynamically balanced case.  E-I perturbations are somewhat amplified for the parametrically balanced case (Fig.\ \ref{fig:balance_perturbations}b, bottom). For these studies, we examined the effect not merely on the output, as in Fig.\ \ref{fig:balance_perturbations}a, but rather on the full network activity, defining
\begin{equation}
\Delta_x = \frac {\int dt \sum_i (x_i(t) - \tilde x_i(t))^2} {\int dt \sum_i (x_i(t))^2} \, ,
\label{eqnDelta}
\end{equation}
where $x(t)$ is the noiseless activity of the rate network and $\tilde x(t)$ the perturbed activity.  We expect similar results to hold for spiking networks \cite{HarishHanselAsynchronous}.

\subsubsection*{Autonomous activity in trained networks}

\begin{figure}
\centering{}\includegraphics[width=0.9\textwidth]{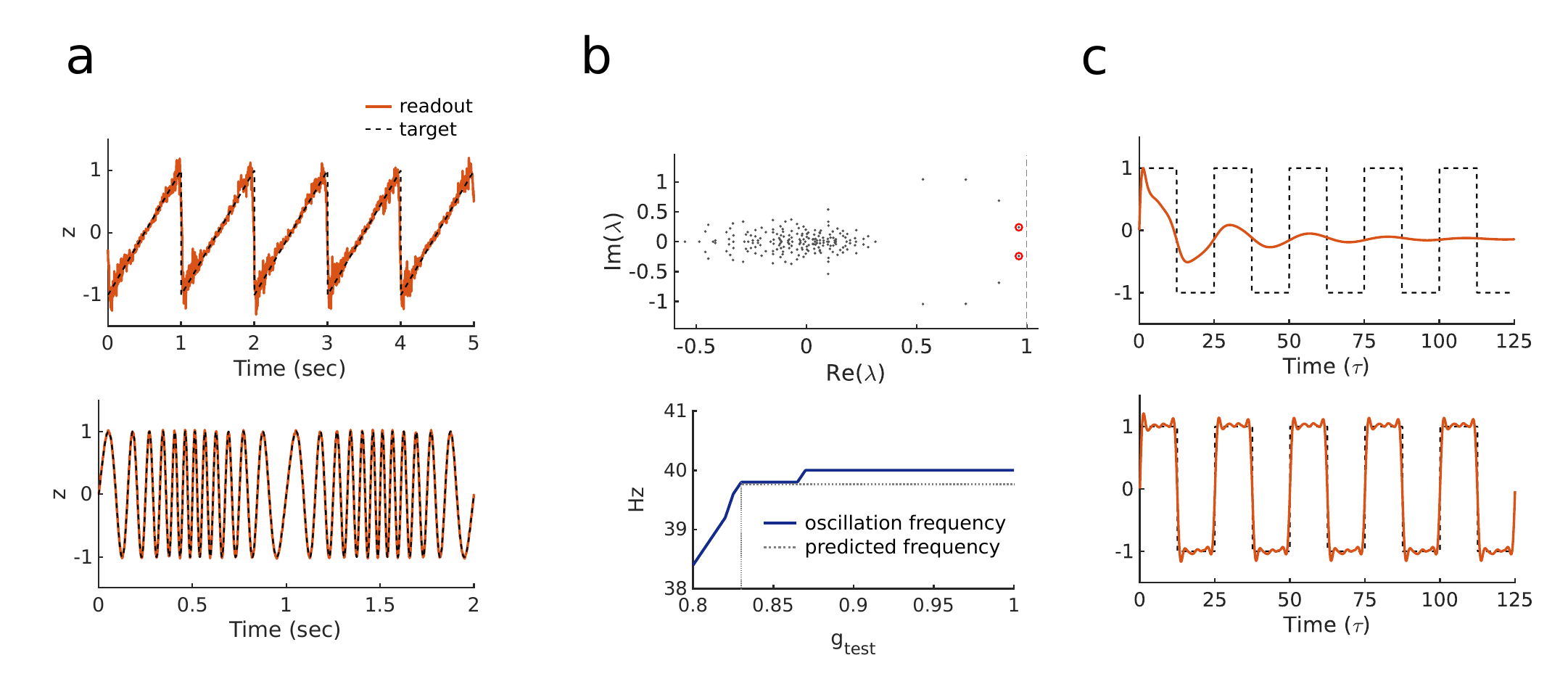}\caption{\label{fig:osc_eig}\emph{Nonlinear oscillations.} \textbf{a})
Top: Balanced E/I spiking network of size $N=300$ producing a sawtooth
wave of frequency $1$ Hz. Bottom: E/I rate network producing a frequency-modulated
oscillation obtained by $F^{out}\left(t\right)=\sin\left(\omega\left( t\right)t\right)$
with $\omega\left(t\right)$ linearly increasing from $2\pi$ to $6\pi$
Hz for the first half of the oscillation period, then reflected in
time around the midpoint of the period. Parameters: $N=500$, $\phi=$halftanh,
trained using feedback (Methods, $\Delta t_{L}=1$ s). \textbf{b}) Top:
Eigenvalue spectrum of $J_{test} \phi'|_{x0}$ for a dynamically balanced rate network
with sigmoid activation function trained to produce a square
wave ($N=200$, output frequency $f=0.04$, $\tau=1$), for $g_{\test}=0.8$. The two red dots indicate the two conjugate eigenvalues $\lambda_{1,2}$ with largest real value. Bottom:  Oscillation frequency as a function of $g_{\test}$ comparing simulation results (solid curve) with approximate prediction (dashed lines). \textbf{c})
Readout signal with $g_{\test} = 0.8$ (top) and $g_{\test}=1.0$ (bottom).}
\end{figure}

We found that the generation of oscillatory activity in trained network (such as that shown in Fig. \ref{fig:osc_eig}a) could be described by a simple mechanism, at least when a single frequency dominates that output pattern. After training, the spectrum of the synaptic matrix $J$ usually shows a complex conjugate pair of eigenvalues with largest real part. This is not limited to target-based learning methods: we trained networks of different sizes using a variety of activation functions using back propagation through time (either employing stochastic gradient descent or ADAM \cite{KingmaAdam}), and we consistently observed this phenomenon for different target readout signals of various frequencies. For differentiable activation functions, the oscillatory frequency is approximately predicted to be $f=\mbox{Im}(\lambda_1)/2\pi\tau\mbox{Re}(\lambda_1)$, where $\lambda_1$ is one of the two complex eigenvalues with largest real part of the matrix $J\phi'|_{x0}$ (Fig. \ref{fig:osc_eig}b), and $\phi'|_{x0}$ is the derivative of the activation function computed at the (not necessarily zero) fixed point from which the oscillations arise by means of a supercritical Hopf transition.

This analysis can be verified after training is completed by artificially lowering the effective gain of the obtained connectivity matrix $J$ using a fictitious gain parameter $g_{\test}$ in the testing phase, such that $J_{\test}=g_{\test}J$. Nonlinear oscillations arise at the critical value $g_{\test}^{*}$ where the previously stable fixed point loses its stability as the two dominant conjugate eigenvalues cross the imaginary axis (Fig. \ref{fig:osc_eig}c). At the bifurcation, the frequency is controlled by the imaginary part of the dominant eigenvalues and the network dynamics is essentially two-dimensional. As $g_{\test}$ is increased, there is a small change of frequency of the readout signal as nonlinear effects start to grow and other frequencies and harmonics kick in (Fig. \ref{fig:osc_eig}b). This picture is consistent with previous work in random E/I separated rate models \cite{delmolinosynchronization} as well as a recent study of low-rank perturbations to randomly connectivity matrices \cite{mastrogiuseppelinking}.

\begin{figure}
\centering{}\includegraphics[width=0.7\textwidth]{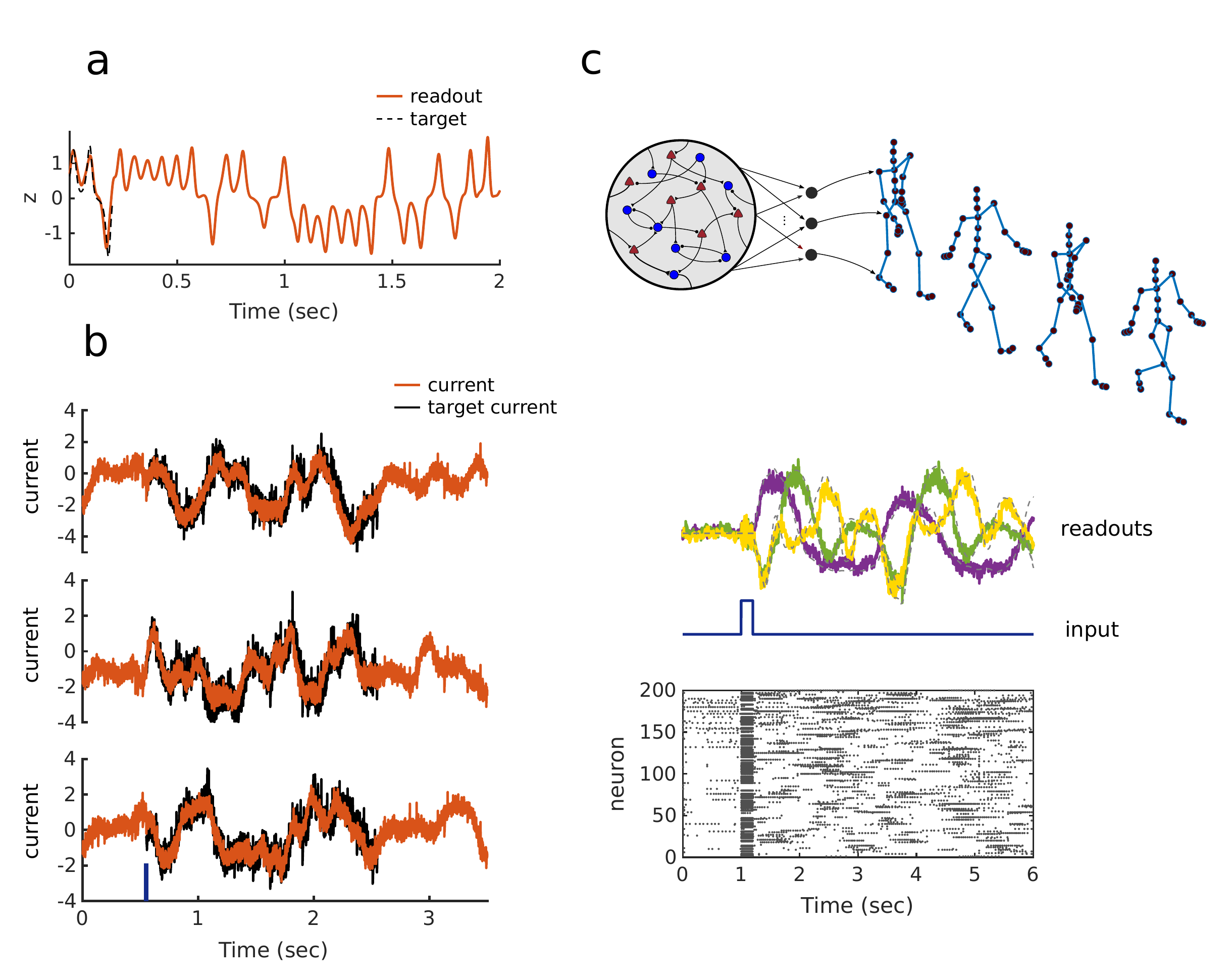}\caption{\emph{\label{fig:learning_complex_transients}Learning chaotic trajectories}
\emph{and complex transient activity}. \textbf{a}) Output of a rate
network ($N=1000$, halftanh activation function) trained
to produce the time course of the first coordinate of a Lorenz attractor
($\sigma=10$, $\rho=28$, $\beta=2.67$).\textbf{ b})
Input currents onto three representative neurons in
a balanced spiking network trained to reproduce innate current trajectories
of duration $2$ s after a brief stimulus ($50$ ms) at time $0.5$
s. Network size $N=500$, synaptic time constant $\tau_{s}=50$ ms.\textbf{
c}) Balanced E/I spiking network producing walking behavior in response
to a strong input pulse of duration $100$ ms. Top:
a pictorial representation of the network with $56$ distinct readouts
(network size $N=300$; synaptic time constant $\tau_{s}=50$ ms).
Middle: activity of three random readout units over the course of
$\sim6$ s. Bottom: spike raster plot of $200$ neurons in the network\textbf{.}}
\end{figure}

Balanced networks can also be trained to produce prescribed chaotic
dynamics (like the Lorenz attractor in Fig.\ \ref{fig:learning_complex_transients}a) or multiple complex quasi-periodic trajectories.  In another task, inspired by the work of Laje, and Buonomano \cite{LajeBuonomanoTamingChaos}
in rate networks, and similar to recent extensions to the
QIF spiking case in \cite{kimlearning}, we trained a spiking network
to reproduce a desired transient dynamics in response to an external
stimulus. To do so, we recorded innate current trajectories $h_{i}^{\T}\left(t\right)$
generated by a randomly initialized LIF balanced network for a short
period of time ($2$ sec) during its spontaneous activity. We then
trained the same network to reproduce its innate current trajectories
whenever a strong external input was applied (dark blue line in Fig.\ \ref{fig:learning_complex_transients}b). The brief external pulse ($50$ ms) is able to elicit the target trajectory, after which the network naturally resumes its
irregular activity.  Finally,  the example in Fig.\ \ref{fig:learning_complex_transients}c shows an
E/I spiking network instructed to generate the quasi-periodic dynamics
of human walking behavior shortly after a $50$ ms unitary pulse.
We trained $56$ linear decoders on the network activity to reproduce
the time-course of each joint-angle from a human Motion-Capture dataset,
as in \cite{sussillo2009generating,alirezaarxiv}. The average firing
rate of the network is $20$ Hz. A brief input pulse can trigger the
motion generation from asynchronous spontaneous activity or
reset the phase of a previously stable quasi-periodic dynamics.

\subsubsection*{Input-Output tasks}

\begin{figure}
\centering{}\includegraphics[width=0.6\textwidth]{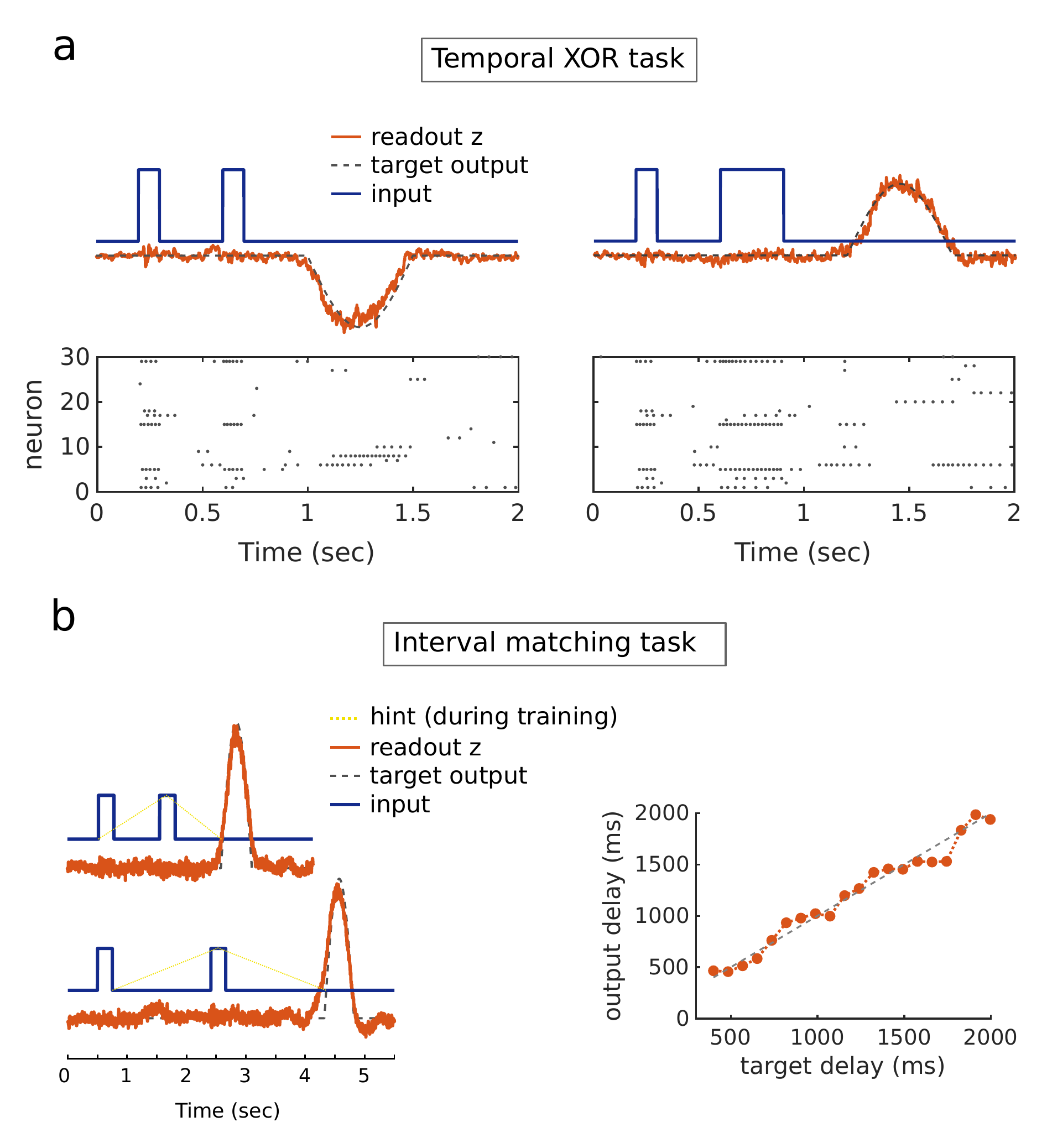}\caption{\label{fig:xor_and_time_matching}\emph{Input-output tasks}. \textbf{a})
Example of output responses (red curves) of a balanced E/I spiking network
trained on the temporal XOR task to two sets of input pulses (Blue
curves) respectively coding for False-False (left) and False-True
(right) Parameters: $N=1000$, $\tau_{s}=50$ ms. \textbf{b}) Interval matching task. Left: sample output (red curves)
vs desired output (dashed black curves) from a spiking E/I network trained on
the Interval Matching Task to two pairs of input pulses. Right: output
delay vs target delay $\Delta T$ to randomly interleaved test input
pulses.}
\end{figure}

Our learning procedure can also be employed to train dynamically balanced E/I networks capable of
performing complex temporal categorization tasks. As our first example,
a spiking network implements an exclusive OR function
\cite{depasquale2016using} anytime an appropriate sequence of inputs
is presented, despite disturbance induced by its spontaneous asynchronous
activity (Fig.\ \ref{fig:xor_and_time_matching}a). In each
trial, the network is presented with two pulses of durations that are
chosen randomly to be either short ($100$ ms) or long ($300$ ms),
coding for the truth values $0$ (False) or $1$ (True).
The network computes the XOR function of the two
inputs and responds with an appropriate positive or negative readout
signal (duration: $500$ ms) after a delay period ($300$ ms). We
used online BCD to train a balanced network of $N=1000$ LIF neurons
and measured the number of correct responses. The trained network responds promptly when
the two impulses are presented at any random time over the course
of its spontaneous activity and reaches a test accuracy of $96\%$.

As a second example, we construct an E/I spiking network to solve
a more complex interval time-matching task, inspired by the ``ready-set-go\textquotedblright{}
task employed in \cite{Jazayeri2010}.  This task has been solved previously
using newtorks with unconstrained synaptic weights \cite{DePasqualeFullForce}. In this task, the network
receives two brief input pulses separated by a random delay $\Delta T$,
and it is trained to generate a response after exactly the same delay,
following the second pulse. As in the temporal XOR task described
above, it is crucial here that the network retains information about
the first pulse during the whole delay period in the absence of any
external input. Expecially for long delays $\Delta T$, this task
proves hard to solve. We therefore employ the heuristic technique
of ``hints'' previously introduced in \cite{DePasqualeFullForce}:
in each training epoch, the teacher network is provided with both
a ramping up and decreasing input (dashed yellow line in Fig.\ \ref{fig:xor_and_time_matching}b, left) during the two relevant delay periods. An E/I network
of $N=1000$ spiking neurons produces accurate responses to random delays
between $400$ ms and $2$ s  (Fig.\ \ref{fig:xor_and_time_matching}b, right).

\section*{DISCUSSION}

We have introduced a fast alternative to RLS that is capable of training sign-constrained
rate-based and spiking network models and, in addition, has the promising features of good
memory and computational requirements when dealing with E/I (and also sparse) models.
We have shown that this fast target-based learning scheme can be used to
train balanced networks of rate and spiking neurons for a wide variety
of tasks. We described the conditions under which dynamically balanced networks
can be obtained with the training procedure. We found that, in the
absence of proper initialization and regularization, learning dynamics
is attracted to regions of weight space with parametrically tuned connectivity,
and we showed the impact of specific weight regularizations on the weight
structure of trained networks, as well as their resilience to various
external perturbations.

\subsubsection*{Relation to other work}

We have tackled the problem of training spiking neural networks to
display prescribed stable dynamics or to solve cognitively relevant
input-output tasks. A number of top-down approaches have been proposed
to train functional models of spiking networks, e.g.\ the neural engineering
framework \cite{EliasmithNeuralEngineering}, spike-coding \cite{BoerlinDeneveSpikeBasedPopulationCoding}
and nonlinear optimal control \cite{DeneveBrainEfficient,alirezaarxiv}.
These methods are elegantly formulated and effective.  Interestingly, they solve a different task than what our procedure solves.  These methods train the network to reproduce a prescribed dynamics, whereas our method trains a network to produce a particular trajectory generated by those dynamics.  The resulting two networks look identical as long as the prescribed trajectory is being followed, but they generalize differently if the network deviates from this trajectory.

Some variations of RLS-based training have been introduced previously
to construct functional models of E/I separated spiking networks.
In \cite{NicolaClopathSupervised}, the authors employed a clipping
procedure on top of a FORCE training method, which entails rank-1
updates to the original randomly connected recurrent network, while
in \cite{depasquale2016using} the authors used an off-line
two step Full-FORCE procedure to train a large network performing
an oscillation task. In a slightly different setting, the authors of \cite{kimlearning} used Full-FORCE to train networks of quadratic integrate and fire neurons to reproduce prescribed
synaptic drive, as well as spiking rate patterns in response to a
brief strong stimulus. They provide an example of an E/I network with
parametrically tuned effective connectivity and no external currents that tracks
its own innate trajectories, recorded over the course of spontaneous
activity. Sign constraints were imposed by eliminating updates of synapses
that would pass out of the allowed ranges in a given epoch, and those synapses
were then deleted in subsequent epochs (we call this strategy Clipped-RLS).
Although performance of Clipped-RLS is comparable to BCD, this strategy
proves memory-demanding for large network sizes, especially when dealing
with dense topologies. Clipped-RLS entails using $N$ independent
covariance matrices $P_{i}$, one for each unit in the trained network,
thus amounting to storing $N\times\left(pN\right)^{2}$ floating-point numbers
(FPs). For comparison, BCD requires $2N^{2}$.

\subsubsection*{Conclusions}

Credit-assignment is a major problem in training spiking networks,
where differentiability issues limit  the use of gradient-based optimization (but see \cite{LeeBP,HuhBP,ZenkeBP}), which has proven very powerful in deep feed-forward architectures. Whereas in some approaches the credit assignment problem is tackled by relying on coding assumptions variably linked to optimality criteria, target-based approaches, both in the context of feed-forward \cite{LeeTargetPropagation}
and recurrent models, provide a straightforward solution. As shown
above as well as in a recent work \cite{kimlearning}, it is not essential
for the teacher network to be a rate model, as long as it effectively
acts as a dynamic reservoir that expands task dimensionality via
its recurrency, therefore proving rich targets.

\section{Methods}

\subsection{Rate and spiking networks models}

The weight matrix $J$ is initialized by setting $J_{ij}= J_{\X\Y}^{\eff}/\sqrt{N_{\preM}}+\Delta_{ij}$,
where X and Y are the appropriate E and I labels corresponding to neurons $i$ and $j$. $\Delta_{ij}$ is a random matrix with entries that are zero-mean Gaussian distributed with each column $j$ having variance $g^{2}/N_{\preM}$. When a balanced teacher network
is employed during training, we use a non-negative activation function
and appropriately choose block averages and external constant currents $I\propto\sqrt{N}$ for which the balance equation yields a solution with appreciable positive rates.  In those cases where we seek to train spiking networks displaying irregular spontaneous activity with low rates, we further adjusted
the random part $\Delta_{ij}$ so that $\sum_{j}\Delta_{ij}=0$ for
each row $i$.  By reducing quenched fluctuations in time-averaged
activities for each neuron, this method ensures that spiking neurons
trained on the teacher currents do not have abnormally low or large
average activity.

Integration of ODEs is performed by the forward Euler method using an
integration time-step not larger than $\Delta t= \tau/20$
for rate models and $\Delta t=0.5$ ms for spiking networks. We further scale down the integration time-step
in all those case where large $J_{\X\Y}^{\eff}$ and strong external currents are employed.

\subsection{Learning algorithm}

\paragraph*{Bounded Coordinate Descent}

When training a rate or a spiking network, we seek to match the incoming
currents in the driven teacher $h_{i}^{\T}\left(t\right)=\sum_{j}J_{ij}^{\T}\phi\left(x_{j}^{\T}\left(t\right)\right)+\sum_{k}w_{ik}^{\out}F_{k}^{\out}\left(t\right)+I_{i}$
with those in the student: $h_{i}\left(t\right)=\sum_{j}J_{ij}\phi\left(x_{j}\left(t\right)\right)+I_{i}$
(for a rate student) or $h_{i}\left(t\right)=\sum_{j}J_{ij}s_{j}\left(t\right)+I_{i}$ (for a spiking student).
In training spiking networks, performance is virtually
unchanged if one were to choose to match the activity $x^{T}\left(t\right)$
in the teacher rate network with the synaptic currents $h\left(t\right)=Js\left(t\right)+I$
in the spiking network. We sometimes allow for an additional scaling and/or
offset of the currents provided by the teacher network, so that the
actual target currents are defined as $\omega_{h}h_{i}^{T}\left(t\right)+b_{h}$. Each neuron is trained independently and in parallel every $\Delta t_{l}$, after a transient $T_{d}$ to wash out the initial condition.

We optimize the loss-function with an online strategy by means of
Bounded Coordinate Descent (BCD). In our case, the method consists
in updating, in parallel for each postsynaptic neuron $i$, each
synapse $J_{ij}$ one at a time by computing the optimal solution
to the one-dimensional optimization problem where all other synapses
$J_{ik}$ for $k\neq j$ are kept fixed: 
\begin{equation}
J_{ij}\to\frac{J_{ij}C_{ii}+\alpha tW_{ij}+D_{ij}}{C_{ii}+\alpha t}\label{eq:coordinate_descent}
\end{equation}
where $C$ is the covariance matrix of the activities $C_{ij}\left(t\right)=\sum_{\tau=0}^{t}s_{i}\left(\tau\right)s_{j}\left(\tau\right)$, which gets updated at each time-step by $C_{ij}\to C_{ij}+s_{i}s{}_{j}$ (these equations are for the spiking case; for rate models $s_i$ is replaced by $\phi(x_i)$).
The residual matrix $D_{ij}$ is defined as $D_{ij}\left(t\right)=\sum_{\tau=0}^{t}s_{j}\left(\tau\right)\left(h_{i}^{\T}\left(\tau\right)-h_{i}\left(\tau\right)\right)$.
After each update with change $\Delta J_{ij}$, the $i$th
row $D_{i:}$ of the residual matrix $D$ gets updated according to
\begin{equation}
D_{i:}\to D_{i:}-\Delta J_{ij}C_{j:}\label{eq:residual_update}
\end{equation}
where $C_{j:}$ stands for the $j$th row of $C$. Setting $W_{ij}=J_{ij}^{0}$, where $J^{0}$ is the initial weight matrix, we implement the J0 regularizer. Alternatively, $W_{ij}=0$ corresponds to a simple L$2$ weight regularization.

The updating schedule of weight indexes $j\in\left\{ 1,2,...N\right\} $ can be either fixed
or random at every step. For easier tasks, updating a random subset
of incoming synapses at each time-step is enough to obtain good
training performance. We do not update the weights when this would
violate the imposed sign constraints.

One of the benefits of BCD, compared to local optimization approaches
(e.g.\ stochastic gradient decent), is its ability to keep the neural trajectory close to
the target during training, even more so in the presence of strong external currents that prevent the network from shutting-down.

We note that coordinate descent proves a versatile method
even beyond the sign-constraind case. For example, in updating incoming synapses
to neuron $i$, it is easy to account for specific network topologies
of the $J$ matrix by selecting a relevant subset of rows/colums of
the (symmetric) matrix $C$ in the update equation \eqref{eq:coordinate_descent}.

\paragraph*{Regularization}

In addition to the regularizations discussed in the text, we also experimented with a regularization of the form \begin{equation}
\sum_{\X\in\left\{ E,I\right\} ,j\in\X}\left(J_{ij}-\frac{\sum_{k\in\X}J_{ik}}{N_{\X}}\right)^{2} \, ,
\end{equation}
which controls the variance of the outgoing synaptic weights in each
sub-population. For simple tasks, this typically produces inhibitory
dominated networks with a non-singular $J^{\eff}$.

\paragraph*{Feedback stabilization}

In some cases, it is useful to use a feedback mechanism to
speed-up training and drastically reduce the frequency of weight update
$1/\Delta t_{l}$.  Specifically, during training we drive the
student network with a modified current $\tilde{h}_{i}=h_{i}+\kappa\left(t\right)\left(h_{i}^{\T}-h_{i}\right)$.
We use $\kappa\left(t\right)=\left|h-h^{\T}\right|/(\left|h\right|+\left|h^{\T}\right|)$,
with $|h|$ the Euclidian norm of the vector $h$ (although good training
performance can be achieved with different metrics). The choice of
an adaptive-gain feedback procedure frees from hyper-parameter optimization
of the time-course of $\kappa\left(t\right)$, which is usually taken
to be a decreasing function of time. It is also instrumental in providing
a minimal supervisory signal, thus allowing the student network to
progressively exploit its own fluctuations over the course of training
to build stability around the target trajectory. More generally, other
kinds of constrained optimization methods, e.g.\ interior point methods,
tend to work well when coupled with the feedback mechanism.

\paragraph*{Testing}

Test error is computed over a testing period $T_{test}$ as 
\begin{equation}
E_{\test}=\frac{\sum_{k=1}^{K_{\out}}\sum_{t=0}^{T_{\test}}\left(z_{k}\left(t\right)-F_{k}^{\out}\left(t\right)\right)^{2}}{\sum_{k=1}^{K_{\out}}\sum_{t=0}^{T_{\test}}\left(F_{k}^{\out}\left(t\right)\right)^{2}} \, .
\end{equation}
For input-output tasks, we randomly initialize the network state at
the beginning of a test trial. For periodic targets $F^{\out}\left(t\right)$,
testing is interleaved with training, so that the spiking (rate) network
state $\boldsymbol{s}$ ($\boldsymbol{x}$) is usually close to the
target trajectory. In this case, a sufficiently low test error usually
implies the presence of a stable limit cycle, and the periodic output
is reproduced, up to a phase shift, starting from any initial condition.

For the XOR task,  during testing we defined a correct response when the normalized dot product of the readout $z$ and $F^{\out}$, with $t$ in the window of non-zero target, satisfied
\begin{equation}
\frac{\sum_{t} z\left(t\right)F^{\out}\left(t\right)}{\sqrt{\sum_{t} z^{2}\left(t\right)}\sqrt{\sum_{t}\left(F^{\out}\left(t\right)\right)^{2}}} > 0.5 \, . 
\end{equation}

\begin{acknowledgments}
We thank Laureline Logiaco, Fabio Stefanini and R. Engelken for fruitful discussions.  Research supported by NSF NeuroNex Award DBI-1707398, the Gatsby Charitable Foundation, and the Simons Collaboration for the Global Brain. 
\end{acknowledgments}

 \bibliographystyle{unsrt}
\bibliography{references}

\end{document}